\documentclass[12pt,preprint]{aastex}

\usepackage{natbib}
\usepackage{enumerate}
\usepackage{enumitem}

\slugcomment{}

\shorttitle{Tether-cutting Reconnectin During a Major Solar Event}
\shortauthors{Chen et al.}

\begin{document}


%\title{EUV Arch preventing eruption of a filament , \\
%    Gauge-Boson Couplings, and AAS\TeX\ Examples}
%\title{Overlying Extreme-ultraviolet Arcades Preventing Eruption of a
%        Filament Observed by AIA/SDO}
%\title{Direct Observations of Tether-cutting Reconnection During a Major
 %       Solar Event on 2014 February 25}
\title{Direct Observations of Tether-cutting Reconnection During a Major 
        Solar Event From 2014 February 24 to 25}
%\title{EUV Observations of T-C reconnection During a Major Solar Event by SDO/AIA}

\author{Huadong Chen\altaffilmark{1}, Jun Zhang\altaffilmark{1}, Xin Cheng\altaffilmark{2}, 
Suli Ma\altaffilmark{3}, Shuhong Yang\altaffilmark{1} and Ting Li\altaffilmark{1}}
\email{hdchen@nao.cas.cn}
%\affil{College of Science, China University of Petroleum, Qingdao 266580, China}
%\affil{Key Laboratory of Solar Activity, National Astronomical Observatories, Chinese Academy of Sciences, Beijing 100012, China}
\altaffiltext{1}{Key Laboratory of Solar Activity,
     National Astronomical Observatories, Chinese Academy of Sciences,
      Beijing 100012, China}
\altaffiltext{2}{School of Astronomy and Space Science, Nanjing University, Nanjing 210093, China}
\altaffiltext{3}{College of Science, China University of Petroleum, Qingdao 266580, China}
%\author{S. Djorgovski\altaffilmark{1,2,3} and Ivan R. King\altaffilmark{1}}
%\affil{Astronomy Department, University of California,
%    Berkeley, CA 94720}

%\author{C. D. Biemesderfer\altaffilmark{4,5}}
%\affil{National Optical Astronomy Observatories, Tucson, AZ 85719}
%\email{aastex-help@aas.org}

%\and

%\author{R. J. Hanisch\altaffilmark{5}}
%\affil{Space Telescope Science Institute, Baltimore, MD 21218}

%\altaffiltext{2}{Society of Fellows, Harvard University.}
%\altaffiltext{3}{present address: Center for Astrophysics,
%    60 Garden Street, Cambridge, MA 02138}
%\altaffiltext{4}{Visiting Programmer, Space Telescope Science Institute}
%\altaffiltext{5}{Patron, Alonso's Bar and Grill}


\begin{abstract}
Using the multi-wavelength data from Atmospheric Imaging Assembly on board the Solar 
Dynamic Observatory, we investigated two successive solar flares, a C5.1 confined 
flare and an X4.9 ejective flare with a halo coronal mass ejection, in NOAA AR 11990 from
2014 Feb 24 to 25.
Before the confined flare onset, EUV brightening beneath the filament was detected. 
As the flare began, a twisted helical flux rope (FR) wrapping around the filament
moved upward and then stopped, and in the meantime an obvious X-ray source 
below it was observed.
%, rose with a projected velocity of $\sim$67.8 km s$^{-1}$ and finally stopped at a height of $\sim$19 Mm.
%An obvious X-ray source in the energy range of 10--20 keV appeared below the helical flux rope.
 %suggesting magnetic reconnection is very likely to occur there.
%By means of differential emission measure (DEM) analysis, we find that the newly-formed flux rope had a mean density of $\sim$8.0 $\times$ 10$^{9}$ cm$^{-3}$ and an average temperature of $\sim$10 MK.
%In the AIA 94 \AA\ channel, 
Prior to the ejective X4.9 flare, some pre-existing loop structures in the active region
interacted with each other, which produced a brightening region beneath the filament.
Meanwhile, a small flaring loop appeared below the interaction region
and some new helical lines connecting the far ends of the 
loop structures was gradually formed and continually added into the former twisted FR.
%and a large-scale twisted flux rope connecting the far ends of the loop structures was gradually formed.
%At the onset of the eruption, the newly-formed flux rope apparently imposed an upward tension force to the filament and drove it to erupt outward.
Then, due to the resulting imbalance between the magnetic pressure and tension, the new FR together with the filament erupted outward.
%At the onset of the eruption, the filament was ejected outward by the new large-scale flux rope due to magnetic tension.
Our observations coincide well with tether-cutting model, suggesting that the two flares probably have the same triggering mechanism, i.e., tether-cutting reconnection.
%The detailed process of tether-cutting reconnection is clearly exhibited by our observations.
To our knowledge, this is the first direct observation of tether-cutting reconnection occurring 
between the pre-existing loops in active region.
In the ejective flare case, the erupting filament exhibited an $\Omega$-like kinked structure and underwent an exponential rise after a slow-rise phase, indicating the kink instability might be also responsible for the eruption initiation.
%We simply discussed the possible relationship between the total twist stored in the filament and the special $\Omega$-like erupting structure.

%An obvious X-ray source in the energy range of 10$\sbond$20 keV between the newly-formed flux rope and the lower small flaring loop was detected by the RHESSI observations, indicating the likely occurrence of magnetic reconnection.

\end{abstract}


%\keywords{globular clusters: general --- globular clusters: individual(NGC 6397,
%NGC 6624, NGC 7078, Terzan 8}



\section{Introduction}
%Solar filaments are large magnetic structures supporting cool, dense chromospheric material  against solar gravity in the surrounding hot, tenuous corona.
% (called prominences when observed above the solar limb) 
%($\sim$10$^4$ K and 10$^{10}$--10$^{11}$ cm$^{-3}$)
%($\sim$10$^6$ K and 10$^{7}$--10$^{9}$ cm$^{-3}$)
%Usually, they erupt eventually accompanied by flares and coronal mass ejections \citep[CMEs, e.g.,][]{priest02}. At times, the filament eruptions are ``failed'' \citep[e.g.,][]{ji03,liu09,mrozek11,netzel12,chen13}. 
%In this case, no obvious CMEs and EUV waves would be observed in the wake of the eruptions; The related flares are called confined flares. Strong constraining from the overlying magnetic arcades is believed to be an important reason for those failed filament eruptions \citep[e.g.,][]{torok05,shen11,netzel12,chen13,yang14}.

Unraveling the triggering mechanism of solar eruptions has long been a challenge.
%remains a mystery for many years.
A variety of models have been devoted to interpreting the eruption initiations
\citep[as reviewed by, e.g.,][]{lin03,vrsnak08, chen11}.
Considering removing the strong stabilizing force from the overlying magnetic arcade by
means of a slow magnetic reconnection in the low corona,
\citet{moore80} and \citet{moore01} proposed the tether-cutting mechanism based on a single bipolar field geometry. 
In this model, the strongly-sheared core fields are overlaid by less-sheared envelope magnetic arcades.
%a stressed core field receives a strong stabilizing force from the magnetic tension of its overlying magnetic arcade, whose field lines are analogous????? to tethers. 
At first, the strongly-sheared core fields with opposite polarities reconnect slowly
above the polarity inversion line (PIL), but beneath the filament, which leads to the 
formations of large-scale twisted flux rope connecting the far ends of the core
fields and small flaring loop shrinking downward.
%As more and more ``tethers'' are cut, the magnetic pressure 
Then, due to the out-of-balance situation between the outward magnetic pressure and the downward magnetic tension, the whole field expands outward.
%Then, due to magnetic tension, the concave-upward twisted flux rope propels the filament  
%to erupt outward.
%which stretches up the overlying envelope magnetic field are stretched up,
%magnetic neutral line
As the overlying envelope magnetic arcades are stretched up by the erupting filament, 
fast reconnection occurs with an elongated current sheet forming below the filament, which 
produces the flare ribbons and further speeds up the filament eruption to form the coronal
mass ejection (CME).
A similar mechanism was also discussed by \citet{van89} in their ``flux cancellation'' model.

The tether-cutting or flux cancellation model was supported by some observational
and numerical studies \citep[e.g.,][]{wang93,amari99, amari00,
zhang01,sterling05,yurchyshyn06,sterling07pasj,kim08, liu08,green09, 
liu10,liu12} and has been mentioned in a large number of literatures.
However, most of the evidences supportive of tether-cutting reconnection are indirect,
only derived from some associated observational phenomena, such as H$\alpha$, 
EUV, or X-ray brightenings \citep[e.g.,][]{moore80,sterling05,yurchyshyn06}, 
slow-rise motion of filament \citep[e.g.,][]{sterling07pasj,sterling11}, 
morphological changes of flaring structures \citep[e.g.,][]{kim08,liu10}, and photospheric sheared magnetic fluxes \citep[e.g.,][]{moore92,moore01,savcheva12}.
So far, direct observations of the tether-cutting reconnection have been very rare.
The ideal kink instability of twisted magnetic flux ropes is considered as another possible 
initial driver of solar eruptions \citep[e.g.,][]{torok03,kliem04}, which is supported by the frequently observed deformation of the flux rope axis during eruption \citep[e.g.,][]{ji03,williams05,yang12,shen12}. 
%The kink instability sets in when the amount of magnetic twist in the flux ropes exceeds a critical value, likely ranging from 2$\pi$ to 6$\pi$ \citep[e.g.,][]{hood79,torok04,torok14}. In a few cases, the erupting filaments of very high twist (6$\pi$$\sim$10$\pi$) have been reported as well \citep{romano03,gary04,kumar12}.

In this study, using the high-resolution multi-wavelength data from Atmospheric Imaging Assembly/Solar Dynamic Observatory \citep[AIA/$\it{SDO}$;][]{lemen12}, we present an 
unambiguous observation of tether-cutting reconnection during a major solar eruption,
which occurred between the pre-existing loop structures in active region (AR) 11990 and 
most likely triggered the ensuing X4.9 flare and associated halo CME with a combination of 
kink instability of the filament.

\section{Data and Observations}
%\usepackage{enumerate}
%\begin{enumerate}[label=\emph{\roman{*}})]
%\item 
%We mainly used the data from AIA/SDO.
AIA/$\it{SDO}$ provides full-disk images up to 0.5 R$_{\sun}$  above the solar limb with 
1$\arcsec$.2 spatial resolution and 12 s cadence in 10 wavelengths. 
%For our event, some data were discarded due to saturation, which led to an about 24 s cadence during the period of the flare.
%All the ten bandpasses have been employed in the observations of this event.
We mainly used the data (Level 1.5 images) at 7 EUV channels
centered at 304 \AA\ (\ion{He}{2}, 0.05 MK),
171 \AA\ (\ion{Fe}{9}, 0.6 MK), 193 \AA\ (\ion{Fe}{12}, 1.3 MK and \ion{Fe}{24}, 20 MK),
211 \AA\ (\ion{Fe}{14}, 2 MK), 335\AA\ (\ion{Fe}{16}, 2.5 MK), 94 \AA\ (\ion{Fe}{18}, 7 MK), and 131 \AA\ (\ion{Fe}{8}, 0.4 MK and \ion{Fe}{21}, 11 MK).
We de-rotated the AIA data for each of the two flares to two respective times 
(Feb 24 21:30 UT and Feb 25 00:30 UT).
%The AIA data associated with the two flare cases have been performed de-rotation, respectively.
The longitudinal magnetograms and continuum intensity images with 
1$\arcsec$.0 spatial resolution and 45 s cadence from Helioseismic and Magnetic Imager
\citep[HMI;][]{schou12} on $\it{SDO}$ help us to analyze the connectivities of the loop structures in AR 11990.
We also used the Ramaty High Energy Solar Spectroscopic Imager \citep[$\it{RHESSI}$;][]{lin02} X-ray observations to see the evolution of X-ray sources during the flare. 
The accuracy in the alignment between AIA and $\it{RHESSI}$ images is estimated to be about 
3$\arcsec$ \citep{zehnder03}.
%with 2$\arcsec$.3 spatial resolution 
%The accuracy of image alignment is estimated to be 2 arcs.
%The accuracy in the co-alignment was better than 1 arcs.
%This allows us to align the blue-wing image with a white-light image with 
%an accuracy better than 1 arcs.
%at 21:30 UT on February 24 and at 00:30 UT on February 25, respectively.
%All AIA data used in this paper have been performed de-rotation.
%We have performed de-rotation for AIA data used in this paper.
%We have used the RHESSI hard X-ray (HXR) observations to see the evolution of HXR sources during the CME initiation associated with M-class flare. -----kumar12
%\end{enumerate}

%\section{Observations}
\subsection{Overview of the Event} \label{FE}
According to $\it{GOES}$-15 observations, a C5.1 flare took place in NOAA AR 11990
(S12E82) from 21:31 UT on 2014 Feb 24.
The AIA data show that no mass or magnetic structure escaped 
from the solar surface during the weak flare. 
Three hours later, an X4.9 flare occurred from the same location with peak at 00:49 UT on Feb 25, 
accompanied by a filament eruption and a halo CME with a median velocity of 
$\sim$1041 km s$^{-1}$ (see CACTus catalogue, http://sidc.be/cactus).
Our observations cover both events well.
%On 2014 February 25, a filament erupted in NOAA AR 11990 (S12E82).
%which was near the eastern limb (S12E82) of the Sun.
%According to $\it{GOES}$-15 observations, 
%the associated flare is an X4.9 class flare with start, peak and end time at 00:39, 
%00:49, and 01:03 UT, respectively.
%A halo CME with a median velocity of $\sim$1041 km s$^{-1}$ (see CACTus catalogue, 
%http://sidc.be/cactus) accompanied the filament eruption and flare.

%The Computer Aided CME Tracking (CACTUS: Robbrecht & Berghmans 2004) reveals that the flare was associated with a fast CME, with a radial speed larger than 1000 km s?1 .
%Robbrecht, E., & Berghmans, D. 2004, A&A, 425, 1097 ----Aulanier 2012
%also found to relate to the event.
%By checking the observations from COR1 and COR2/SECCHI on spacecraft STEREO~B, 
%a fast CME with a mean projected velocity of above 1500 km s$^{-1}$ was found to relate
%to the event.
%On 2014 February 25, a GOES X4.9-class flare occurred in NOAA AR 11990,
%which was near the eastern limb (S12E77) of the Sun.
%According to GOES-15 observations, the start, peak and end time of the flare 
%is 00:39 UT, 00:49 UT, and 01:03 UT, respectively.
%In addition, the observations of $\it{GOES}$-15 show that 
%a C5.1 flare took place in the same active region (AR) just about 3 hr prior to the 
%X-class flare.
%However, according to the AIA data, no mass or magnetic structure was found to escape 
%from the solar surface during the weak flare. 

\subsection{A Flux Rope Appearing During the Confined Flare}
Figures~1(a)--(e) display the general evolution of the confined flare 
in AIA 94 \AA\ (see animation~1  in the online journal for more details).
At 21:22:25 UT, about 8 min before the start of the flare, an obvious 
brightening B$_{1}$ appeared beneath the filament.
As the flare began and developed, a twisted flux rope FR$_{1}$, which is outlined by 
the helical dotted lines in panels (c) and (d), was observed to wrap around the filament and move upward.
%was formed gradually. 
%across the filament, 
%It wrapped around the filament and moved upward. 
Meanwhile, a small flaring loop FL$_{1}$ arose at almost the same location as B$_{1}$.
From RHESSI observations, an X-ray source in the energy range of 10--20
keV (as shown by the red contours in panel (d)) appeared between FR$_{1}$ and FL$_{1}$ 
during the flare, suggesting magnetic reconnection is very likely occurring there.
%It suggests that magnetic reconnection is very likely to occur there.
%The positions of the filament were pointed out by the black arrows in Figure~1.
The black arrows in panels (c) and (d) point to the filament.
It can be seen that the filament passed through and was supported 
by the dip of FR$_{1}$.
%be twined by FR$_{1}$ in the dip.
From 21:34:22 UT to 21:37:48 UT, the top edge of FR$_{1}$ rose by about 14 Mm, 
which derived a mean upward velocity of $\sim$67.8 km s$^{-1}$. 
Then, its apex stopped at a projected height of $\sim$19 Mm.
The kinetics of FR$_{1}$ is exhibited by the time-slit map (panel (f)), 
which is from the AIA 94 \AA\ images along the slit in panel (d).
%In the AIA 171 \AA\ images (see the animation~1see animation 3 in the online journal for more details). ), 
Panel (e) shows the complex loop structures of AR 11990 after the confined
flare. 
Besides the twisted flux rope FR$_{1}$, it appears that some other loop structures,
such as ML$_{1}$ and ML$_{2}$, existed in the AR.
%The time-slit map (panel (f)) from the AIA 94 \AA images along the slit in panel (d)
%displays the kinetics of FR$_{1}$ well.

Applying the differential emission measure (DEM) method \citep[see][for more details]{cheng12}
to the AIA simultaneous data in 6 EUV wavebands, 
we calculated the emission measure (EM) and the temperature of AR 11990 
at the peak of the confined flare, which are displayed in Figures~1(g) and (h), respectively.
%panels (g) and (h) of Figure~1, respectively.
According to the relationship between EM and density \textit{n},
\begin{equation}
n=\sqrt{\textup{EM}/l}
\end{equation}
we estimated the density of FR$_{1}$.
Here, we assume that the depths of FR$_{1}$ along the line of sight \textit{l} 
are approximately equal to its widths w$_{1}$ and w$_{2}$, which are indicated 
by the solid lines in panel (g).
As a result, an almost same mean value of $\sim$8.0 $\times$ 10$^{9}$ cm$^{-3}$ of 
the density is outputted from the different widths (w$_{1}$$\sim$9840 km and
w$_{2}$$\sim$4570 km) and the corresponding mean EMs (EM$_{1}$$\sim$6.3 $\times$ 
10$^{28}$ cm$^{-5}$ and EM$_{2}$$\sim$3.0 $\times$ 10$^{28}$ cm$^{-5}$) along
the solid lines.
According to the temperature map, the plasma in the main stem of FR$_{1}$
has an average temperature of T$_{1}$ $\sim$ 10 MK. 
However, the average plasma temperature T$_{2}$ in the dip region 
is only $\sim$ 2 MK, which may be related to the cool filament material
in the region wrapped by the twisted FR$_{1}$.

\subsection{Tether-cutting Reconnection Triggering the Ejective X4.9 Flare}
As mentioned above, about 3 hr after the confined flare, the filament 
erupted fully from AR 11990 with a strong X4.9 class flare and a halo CME.
During the interval, the loop structures in AR 11990
underwent a series of evolutions and changes, and
the most remarkable observational feature is the tether-cutting
reconnection which occurred between the pre-existing loop structures in the AR.
In Figure~2, we provide the distinct evidences for the tether-cutting reconnection during the major 
solar eruption event.

The AIA 304 \AA\ (panels (a)--(c)) and 94 \AA\ (panels (d)--(f)) images
in Figure~2 display the evolutions of the filament and the associated magnetic loops 
(ML$_{1}$, ML$_{2}$, ML$_{3}$, and ML$_{4}$) before the flare.
%(labeled with``ML$_{1}$'', ``ML$_{2}$'', and ``ML$_{3}$'' in panel (d) and ``ML$_{4}$'' in panel (e)) before the flare.
Figure~2(g) is the same AIA 94 \AA\ image as Figure~1(e), 
but with a larger field of view (FOV).
%, showing ML$_{1}$, ML$_{2}$, and ML$_{3}$ more clearly.
%It shows that not only ML$_{1}$, ML$_{2}$ but also ML$_{3}$ were the pre-existing magnetic loop structures in AR 11990 before the ejective eruption.
In the 94 \AA\ channel, it can be clearly seen that 
%evident 
ML$_{1}$ obviously interacted with ML$_{2}$ since $\sim$23:30 UT, 
%about 1 hr before the flare, 
which produced a brightening region (indicated by the red arrows) beneath the filament.
%in Figure~2(d)\sbond(f)
%Using the same DEM method as above, the average plasma temperature 
%and density in this region are estimated as $\sim$5.5 MK and 
%$\sim$6.2 $\times$ 10$^{9}$ cm$^{-3}$, respectively.
%The black box in panel (f) indicates the area where we calculated the mean values of
%the temperature and EM.
% ($\sim$1.5 $\times$ 10$^{28}$ cm$^{-5}$). 
%The depth of the region along the line of sight is replaced by 
%its width (as shown by the black box), when the density is calculated.
Meantime, a flaring loop FL$_{2}$ (indicated by the yellow arrows) 
appeared just below the interaction region.
We also observed in 304 \AA\ channel the associated brightening B$_{2}$ (panel (a))
underneath the filament and hot mass outflows (as shown by the curved blue
arrows in panel (b)) from the interaction region.
As the interaction between ML$_{1}$ and ML$_{2}$ proceeded, some new helical lines 
connecting the far ends of ML$_{1}$ and ML$_{2}$ were gradually formed and continually
added into FR$_{1}$, which resulted in the twisted flux rope FR$\arcmin$$_{1}$ 
(indicated by the white arrows in panel (i)).
%a new twisted flux rope FR$_{2}$ (indicated by the white arrows in panel(i)) connecting the far ends of ML$_{1}$ and ML$_{2}$ was gradually formed.
Subject to magnetic pressure and tension, FR$\arcmin$$_{1}$ gradually rose up
and expanded with the filament.
%and pushed the filament outward. 
From the movie in 94 \AA\ (animation~2 in the online journal),
we can clearly see that FR$\arcmin$$_{1}$ separated from the interaction region 
at the onset of the eruption and FL$_{2}$ shrank downward simultaneously. 
%Panel (i) shows the erupting filament and FR$_{2}$ during the flare.
%we can clearly see the separating of FR$_{2}$ from the hot interaction region and
%the simultaneous shrinking of FL$_{2}$.
%Through the interaction between ML$_{1}$ and ML$_{2}$, at about 00:42UT, it can
%be clearly seen that a new twisted flux rope FR$_{2}$ (indicated by the white 
%arrows in panel(i)) formed between the far ends of ML$_{1}$ and ML$_{2}$.
%, which rooted in the far ends of ...
All the observations described above are well consistent with the tether-cutting model 
\citep{moore01}.
The interaction between ML$_{1}$ and ML$_{2}$ is very likely a kind of slow
magnetic reconnection, which is similar to the photospheric flux cancellation 
\citep[e.g.,][]{van89,wang93,zhang01}, but occurred in the low corona.
It cut the ``tethers'' constraining the filament and triggered the eruption.
%formed a new twisted flux rope with a larger scale.
%which are like tethers constraining the filament, are cut from being tied to the photosphere, forming a 
Additionally, one AIA 211 \AA\ intensity image was given in panel (h) to display
the fine structure of the erupting filament.
%In comparison with the hot 131 \AA\ channel (panel (i)), the fine structure of the erupting filament is better exhibited in the cool 211 \AA\ channel (see panel (h)).
The obvious spatial difference between the filament and FR$\arcmin$$_{1}$ (panel (i)) can be 
easily discerned.

\subsection{Tether-cutting Reconnection and Kink Instability Resulting in the Filament Eruption} \label{}
%\subsection{Filament and Flux Rope Eruptions} \label{}
According to the multi-wavelength observations from AIA, the 
flux rope FR$\arcmin$$_{1}$ resulting from the tether-cutting reconnection
was only visible as emission in the 94 \AA\ and 131 \AA\ wavebands, which
indicates that the plasma in FR$\arcmin$$_{1}$ was very hot with temperatures as high
as $\sim$7--11 MK.
Figure~3 displays the main eruption process of the cool filament and the hot
FR$\arcmin$$_{1}$ in 131 \AA\ (see animation 3 in the online journal for more details).
From panel (b), it can be seen that FR$\arcmin$$_{1}$ (outlined by the black dashed curve) 
passed under the filament.
At the beginning of the eruption, FR$\arcmin$$_{1}$ seemed to impose an upward driving force 
to the filament.
%During the filament ascended, it was apparently exerted an upward driving force by FR$_{2}$. 
In the meantime, most likely due to magnetic reconnection, 
the filament was partially heated and appeared as both bright (hot) and dark (cool) erupting features,
labeled HF and CF in panel (b), respectively.
As the eruption went on, HF and CF both exhibited obvious twisted and kinked 
structures (see panels (c)--(f)),  which suggests that the kink instability 
may also play a role in the destabilization of the filament.

From Figures~3(e) and (f), we can see that the filament displayed an $\Omega$-like erupting structure,
which is quite distinct from the usual reversed-Y shape kink structure 
\citep[e.g.,][]{ji03,williams05,chen13,cheng14} in morphology and
rarely mentioned in previous studies.
%It is worth pointing out that this unique erupting morphology of filament is distinct from the 
%usual reversed-Y shape kink structure and rarely mentioned in previous studies.
%which is rarely reported by previous studies.
%which is likely caused by the kink instability.
%the $\Omega$-like structure of the erupting filament 
One similar observational case was reported by \citet{romano03} and numerically simulated by 
\citet{torok10} (see Figure 12(b) in their paper).
In their works, a very high magnetic twist ($\sim$10--11$\pi$) stored in the flux rope was proposed to explain the uncommon erupting morphology of filament.
%According to their results, the appearance of the $\Omega$-like structure of the erupting filament
%requires at least $\sim$10 pi of the initial total twist.
For a comparison with their results, we made an estimation.
In panel (f), one helical thread (indicated by the blue dotted line) winding around the other 
bright threads can be clearly identified in the filament.
According to our calculation, the length of the filament axis (indicated by the red line) is 
about 31 Mm when the helix wind around the axis in a circle; The total axis length of the filament
is about 194 Mm. Thus, on the assumption of a uniform twist along the axis of the filament,
a total twist of $\sim$12$\pi$ is derived.
This result is comparable to those reported by \citet{romano03} and \citet{torok10}.
%which is comparable to the results reported by romano 2003 and torok 2010.

%Besides the filament and FR$\arcmin$$_{1}$, there are other loop structures,
%such as ML$_{3}$ and ML$_{4}$ (see Figure~3(b)), involved in this eruption.
Additionally, ML$_{3}$ and ML$_{4}$ (see Figure~3(b)) are also involved in this eruption.
%was another magnetic structure involved in the eruption.
According to the AIA observations, ML$_{4}$ seemed to twine around HF 
and was ejected with the filament finally;
%As for ML$_{3}$, it intersected FR$_{2}$ at the top of the filament;
Like FR$\arcmin$$_{1}$, ML$_{3}$ also passed under the top of the filament 
and erupted with the whole ejective system.
However, due to the projection effect, it is hard to discern the connectivity of ML$_{3}$
during the eruption.
%make sure which magnetic structure ML$_{3}$ connected with during the eruption.

%According to the AIA 131 \AA\ observations, during the filament eruption, ML$_{4}$ appeared to
%contain two bright thinner flux tubes FT$_{1}$ and FT$_{2}$, which are outlined 
%by the purple and blue dotted lines in Figure~3(c), respectively.
%From this panel, it seems that FT$_{1}$ twisted with the filament and passed through its top, 
%while FT$_{2}$ directly went across the upward dip region and connected with 
%the left part of the filament.
%As the filament was ejected outwards, its left part was obviously confined by FT$_{2}$. 
%%exerted a obliquely downward pulling force to the filament. 
%In addition, it can be seen that the filament and FT$_{1}$ exhibited kink structures
%in the area where FT$_{2}$ intersected with them, which are clearly displayed
%by the close-up view in panel (c).
%On the basis of these observations, in our view, the appearance of the $\Omega$-like 
%structure of the erupting filament may be not only related to the driving from 
%FR$_{2}$ but also with the confinement by FT$_{2}$ and the kink instability of 
%the twisted filament itself.
%the twist stored in the magnetic structure of the filament.

To reveal the kinetics of the filament, we made the time-slit map
from the AIA 131 \AA\ images and displayed it in Figure~4(a).
Due to the limited FOV of Figure~3, the open narrow box in 
Figure~3(d) only indicates the lower part of the area where the slit image was made.
Figure~4(a) shows the time profiles of the projected heights 
%(measured starting from the bottom of the opened box) 
of CF and HF.
Note that it is hard to discern the top edge of CF in the 131 \AA\ images
during its rapid rise phase, so the kinetics of CF is not clear after about 00:44 UT.
%after it was impulsively accelerated, so the kinetics of CF is not clear after about 00:44 UT.
%The yellow dotted curve in Figure~4(a) outlines 
%the front of the filament in the time-distance diagram.
%In panel (b), we plotted the time variation of the projected height 
%(measured starting from the bottom of the opened box in Figure~3 (e)) of the filament.
%As a whole, 
It can be seen that the evolution of the filament eruption within the FOV of AIA 
divided into two phases: a slow rise phase and an impulsive acceleration phase.
%We performed linear (00:20:43$\sbond$00:41:38 UT) and second order polynomial 
%(00:42:21$\sbond$00:46:05 UT) fitting to the height of the filament,
We used a function consisting of both a linear and an exponential component and a pure 
exponential function to fit the height-time measurements of CF and HF, respectively. 
The corresponding fitting results are displayed by the blue and red dotted lines in Figure~4(a).
It can be found that the heights of CF and HF coincide well with the fitting curves.
Exponential growth of height profile is believed to be the fundamental kinematic feature
of flux rope eruption due to kink instability \citep[e.g.,][]{torok05}.
In this aspect, our observations also suggest that the kink instability may play a role
in driving the filament eruption.
Time variations of the velocities of CF and HF were derived from the fitting results
and plotted in Figure~4(b).
%According to the fitting results, the velocity-time variations of CF and HF were 
%derived and plotted in Figure~4(b).
According to the velocity curves, CF rose slowly with a mean velocity 
of ~2.7 km s$^{-1}$ prior to the flare, which was very likely caused by the 
tether-cutting reconnection, as reported in previous studies \citep[e.g.,][]{sterling07,sterling11};
%whose early movement is a slow rise (as often occurs; e.g., Tandberg-Hanssen et al. 1980; Kahler et al. 1988; Feynman \& Ruzmaikin 2004), -----sterling 2007
%Filaments often undergo relatively slow rising motions from shortly before the onset of solar eruptions (e.g., Tandberg- Hanssen et al. 1980; Kahler et al. 1988; Chifor et al. 2007; Schmieder et al. 2008; Schrijver et al. 2008; Liewer et al. 2009; Cheng et al. 2010; Xu et al. 2010). ------sterling 2011
As for HF, its velocity increased from tens of km s$^{-1}$ to more than
1200 km s$^{-1}$ during a period of $\sim$6 min, which derives an average
acceleration of $\sim$3.5 km s$^{-2}$.

%We performed linear (00:20:43$\sbond$00:42:22 UT) and exponential 
%(00:43:24$\sbond$00:46:05 UT) fitting to the height of the filament,
%which are respectively displayed by the red and blue dotted lines in panel (b).
%Basically, the observational data are in agreement with the fitting curves.
%It can be found that the observational data are in good agreement with the fitting curves.
%An upward velocity of 3.93 km s$^{-1}$ is derived from the linear fitting results.
%This slow-rise movement of the filament was very likely caused by the T-C reconnection
%prior to the eruption, as reported by previous studies (e.g., ???).
%From about 00:43 UT, the filament began to be impulsively accelerated.
 %During a period of $\sim$3 min, its 
 %was accelerated with an increased velocity from tens of km s?1 to more than 200 km s?1.
 %The slow rise 
%(sterling 2007 apj669 slow rise)
%This kinematic transition from linear to exponential rise 
%A mean upward velocity of 3.39 km s$^{-1}$ and acceleration of 7.19 km s$^{-2}$
%are derived from the linear and second order polynomial fitting results, respectively.

In addition, we calculated the AIA intensity fluxes of the tether-cutting reconnection
region (as indicated by the dotted box in Figure~2(e)) in seven EUV channels.
For convenience, each intensity flux was divided by its initial value. 
In Figure~4(c), the time profiles  of the GOES 1--8 \AA\ flux (dotted line) and
the AIA intensity flux in 94 \AA\ (blue), 304 \AA\ (red), 131 \AA\ (purple), 171 \AA\ (black), 
193 \AA\ (orange), 335 \AA\ (turquoise), and 211 \AA\ (green) are plotted.
It is apparent that the intensity fluxes in all seven AIA channels, especially in the 94 \AA, 
304 \AA\ and 131 \AA\ channels, increased within the period of about 40 min 
before the X-class flare.
The associated process of energy release is very likely to result from the tether-cutting reconnection.
%Furthermore, an interesting phenomenon was found that the time variation of 
%the 94 \AA\ flux is distinct from those of the other six intensity fluxes.
%This may be related to the different ranges of the plasma temperature sensitive to the 
%94 \AA\ and the other six EUV passbands.
%The peak values of the 94 \AA, 304 \AA\ and 131 \AA\ fluxes are about 1.7, 1.6 and 1.5 
%times of their respective initial values.

\section{Summary and Discussion}
We presented the detailed observations of two successive flare cases
occurring in NOAA AR 11990 in this work.
% during 2014 February 24 and 25.
Although one is a C5.1 confined flare and the other is an X4.9 ejective flare with a halo CME,
%the magnitudes of the flares and the final eruption situations of the filament are different,
some similar observational features, such as the preflare brightenings (B$_{1}$ and B$_{2}$), 
the small flaring loops (FL$_{1}$ and FL$_{2}$), and the large-scale twisted flux ropes 
(FR$_{1}$ and FR$\arcmin$$_{1}$), both appeared in the two cases.
%Before the starts of the two flares, brightenings (B$_{1}$ and B$_{2}$) were 
%both observed to appear beneath the filament.
%Subsequently, hot ($\sim$7$\sbond$11 MK) twisted flux ropes (FR$_{1}$ and FR$_{2}$) 
%gradually formed.
%They wrapped and drove the filament to move upwards.
%At meantime, flaring loops (FL$_{1}$ and FL$_{2}$) appeared in the lower 
%atmosphere of the sun just below the reconnection regions.
%of high-temperature and high-pressure. 
All these observations are well consistent with the tether-cutting model
\citep{moore01}, which strongly suggests that the two flares probably
have the same triggering mechanism, i.e., tether-cutting reconnection.
%Although the final eruption situations of the filament in the two flares are different,
%the similar observational features in the two events suggest that they probably
%had the same driving mechanism $\sbond$ i.e. T-C reconnection.
Compared with the other similar observations in previous reports \citep[e.g.,][]{kim08,liu10}, 
our observations clearly exhibit the detailed process of tether-cutting reconnection,
which occurred between the inner legs of pre-existing sheared magnetic loops in the AR.
To our knowledge, this is the first direct observation of tether-cutting reconnection.
According to our results, the reconnection region is most evident
%only visible as emission 
in the AIA 94 \AA\ channel, indicating the temperature of the plasma in the reconnection region is about 7 MK.

To describe the detailed process of the ejective eruption studied here, 
%there is a troublesome problem, i.e. the connectivity of ML$_{3}$ 
we have to think about the connectivity of ML$_{3}$.
%There are several possible scenarios on this issue.
According to the AIA observations, ML$_{3}$ might connect with the filament, 
or/and ML$_{4}$, or with ML$_{1}$ through reconnection during the eruption.
From Figure~2, it seems that the right (northern) end of ML$_{3}$ 
was rooted in the same magnetic field region as that of ML$_{2}$.
Thus, we prefer to believe that the tether-cutting reconnection might also occur between
ML$_{1}$ and ML$_{3}$ before the flare, which produced another large-scale flux rope 
like FR$\arcmin$$_{1}$.
%also driving the filament to erupt outward like FR$\arcmin$$_{1}$.
In Figure~5, we drew a schematic diagram on the background of HMI continuum intensity images 
to display this possible scenario and describe the early phase of the ejective eruption.
%The schematic diagram in Figure~5 displays this possible scenario and describes the early phase of the ejective 
Before the onset of the eruption, the filament was located at the PIL of 
the AR, and overlaid by the sheared fields ML$_{1}$, ML$_{2}$ and ML$_{3}$,
which connected the positive polarity regions P$_{2}$ and P$_{1}$ with the negative 
polarity regions N$_{3}$, N$_{2}$, and N$_{1}$, respectively.
ML$_{4}$ winded around the filament with one end rooted in the positive polarity region P$_{3}$.
%Using the HMI line-of-sight magnetograms and allowing for the projection effect, we have confirmed that the two ends of ML$_{1}$, ML$_{2}$ and ML$_{3}$ were respectively rooted in the positive polarity fields P$_{2}$ and P$_{1}$ and the negative polarity fields N$_{3}$, N$_{2}$, and N$_{1}$ before the onset of the eruption.ML$_{4}$ wound around the filament with one end rooted in the positive polarity fields P$_{3}$.
While the inner legs of ML$_{1}$, ML$_{2}$ and ML$_{3}$ approached each other, the reconnection
occurred between ML$_{1}$ and ML$_{2}$ and between ML$_{1}$ and ML$_{3}$,
resulting in the formation of the flux ropes FR$\arcmin$$_{1}$ (enhanced FR$_{1}$) and FR$_{2}$ and
the small flaring loop FL$_{2}$.
%The symbol ``X'' in panel (b) denotes the position of the reconnection.
As the ``tethers'' (ML$_{1}$, ML$_{2}$ and ML$_{3}$) were cut, the outward magnetic pressure was 
out-of-balance with the downward magnetic tension and then caused the whole field to expand.
%magnetic pressure (which would be out-of-balance with the tension immediately after the "tethers" have been cut) may be responsible for causing the field to expand.
%Then, due to the magnetic tension, FR$\arcmin$$_{1}$ and FR$_{2}$ pulled up the filament to erupt.
Meanwhile, the filament underwent an exponential rise and exhibited an 
$\Omega$-like erupting structure subject to kink instability.

\acknowledgments
%The authors sincerely thank the referee for very helpful and constructive comments that improved this paper.
%We acknowledge the AIA team for the easy access to calibrated data. The AIA data are courtesy of SDO (NASA) and the AIA consortium.
SDO is a mission of NASAs Living With a Star Program.
%The STEREO/SECCHI data are produced by an international consortium: NRL, LMSAL, NASA, GSFC (USA); RAL (UK); MPS (Germany); CSL (Belgium); and IOTA, IAS (France). We also thank the LASCO/SOHO team for the data support.
This work was supported by the National Natural Science Foundation of China
(11025315, 11103090, 41204124, 11303016, 11221063, 11203037, 11303050 and 41331068), the National Basic Research Program of China under grant G2011CB811403, the CAS project KJCX2-EW-T07, the Strategic Priority Research Program--The Emergence of Cosmological Structures of the Chinese Academy of Sciences (No. XDB09000000).

\clearpage

%% the file, uncomment the \plotone, \plottwo, and
%% If you need a layout that cannot be achieved with \plotone or
%% \plottwo, you can invoke the graphicx package directly with the
%% \includegraphics command or use \plotfiddle. For more information,
%% please see the tutorial on "Using Electronic Art with AASTeX" in the
%% documentation section at the AASTeX Web site,
%% http://www.journals.uchicago.edu/AAS/AASTeX.
%%
%% The examples below also include sample markup for submission of
%% supplemental electronic materials. As always, be sure to check
%% the instructions to authors for the journal you are submitting to
%% for specific submissions guidelines as they vary from

%% This example uses \plotone to include an EPS file scaled to
%% 80% of its natural size with \epsscale. Its caption
%% has been written to indicate that additional figure parts will be
%% available in the electronic journal.

\begin{figure}
\epsscale{0.8}
\plotone{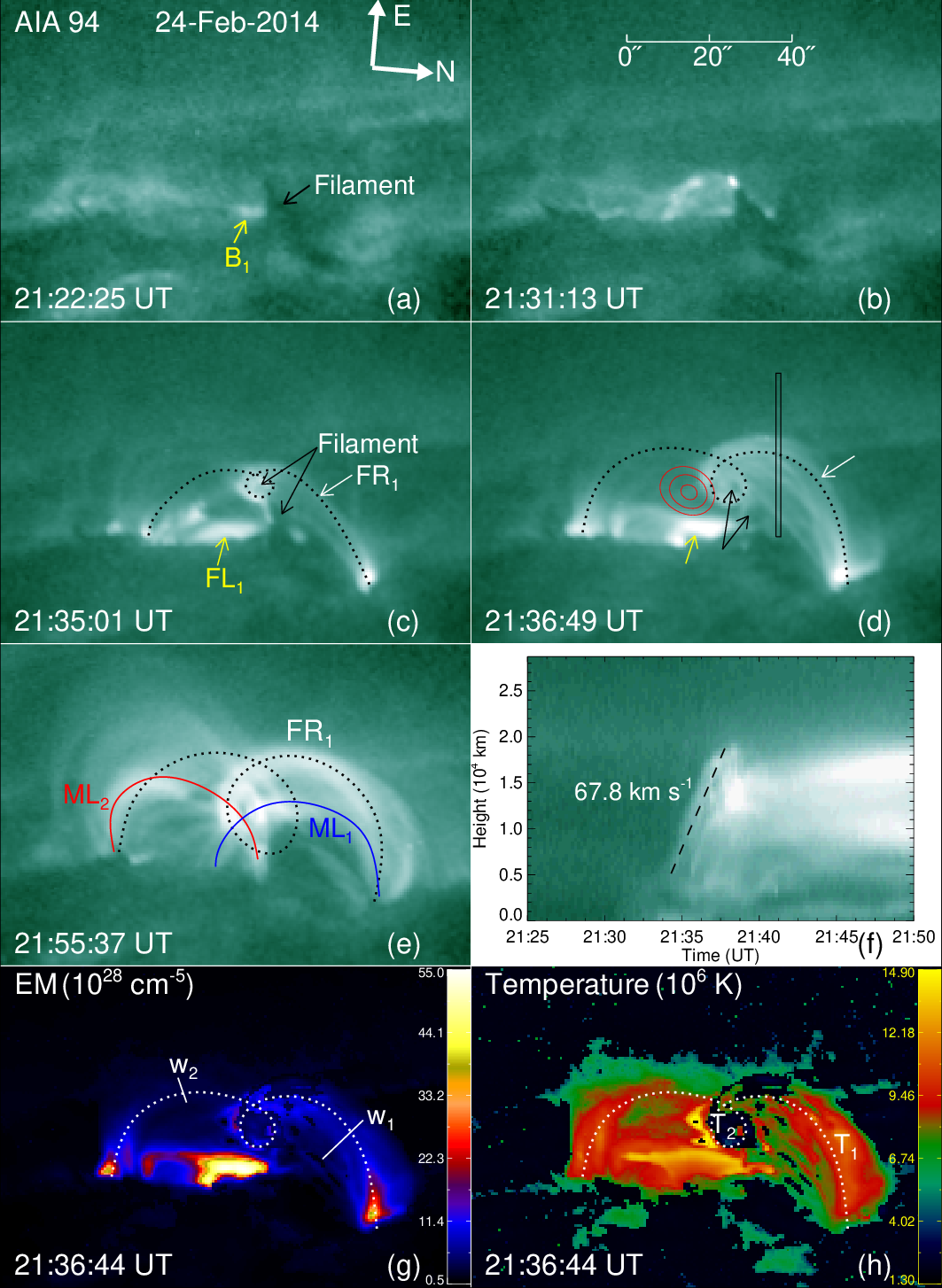}
\caption{(a)--(e) AIA 94 \AA\ images (also see animation~1);
% show the evolution of the confined flare in AR 11990 on 2014 February 24.
(f) Time-slit map from the AIA 94 \AA\ images; 
%along the slit in panel (d). 
(g) EM (h) Temperature map of AR 11990 at the peak of the confined flare.
%The dotted lines indicate the general profiles of the twisted flux rope formed during the confined flare.
In panel (d), the arrows point to the respective features shown in panel (c);
The red contours are from the 20-second 
$\it{RHESSI}$ ``clean'' map in the energy range of 10--20 keV. 
The contour levels correspond to 90\%, 95\%, and 99\% of the maximum value, respectively.
%The levels for the contours are 90\%, 95\%, 99\% of the maximum value, respectively.
The images have been rotated clockwise by 95\degr, which is the same for all AIA and HMI images in the following 
figures.
%in a clockwise direction   ,   from the northern pole of the Sun
The field of view (FOV) is 114\arcsec $\times$ 78\arcsec.
%For detailed descriptions of the labels, see in the text.
%The black slit in panel (d) indicate the wide slit used to make the time-slit maps in Fig.~5.
\label{fig1}}
\end{figure}

%\clearpage

%% Here we use \plottwo to present two versions of the same figure,
%% one in black and white for print the other in RGB color
%% for online presentation. Note that the caption indicates
%% that a color version of the figure will be available online.
%%
\begin{figure}
\epsscale{1.0}
\plotone{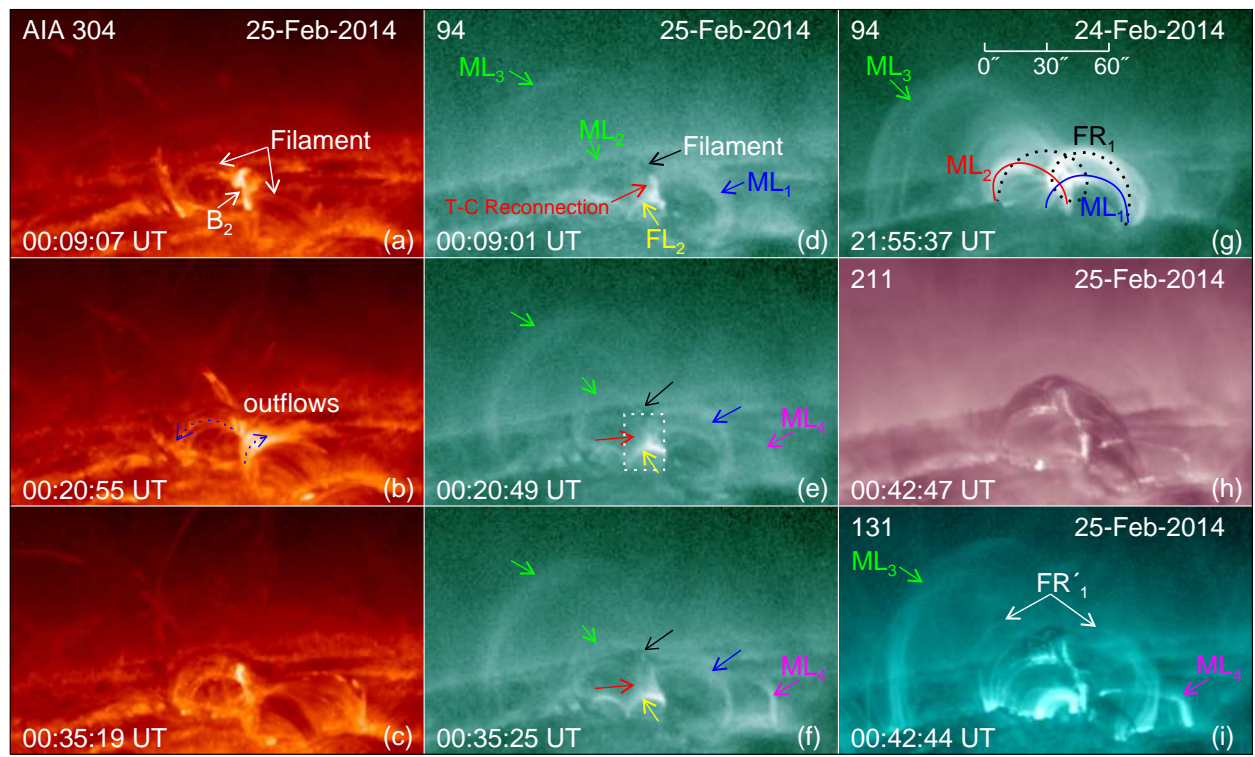}
\caption{(a)--(c) AIA 304 \AA\ and (d)--(f) AIA 94 \AA\ images (also see animation~2) showing the 
tether-cutting reconnection prior to the X4.9 ejective flare;
(g) AIA 94 \AA\ image exhibiting the complex loop structures of AR 11990 
after the confined flare;
(h) AIA 211 \AA\ image showing the fine structure of the erupting filament.
(i) AIA 131 \AA\ image displaying the hot flux rope FR$\arcmin$$_{1}$ formed due to the tether-cutting reconnection.
The unlabelled arrows in panels (e) and (f) are defined in panel (d), 
with arrows of identical colors showing the same respective features.
%the caption should explain that for figures 2d, 2e, and 2f, the arrows are all defined in panel 2d, and the two panels below that panel have the same arrows, with arrows of identical colors showing the same respective features. 
%The blue curved arrows indicate the mass outflows from the T-C reconnection region.
%The dotted box in panel (e) indicates the region where we calculated the AIA intensity flux in Figure~4(c).
The FOV is 200\arcsec $\times$ 120\arcsec.
%For detailed descriptions of the labels, see in the text. 
\label{fig2}}
\end{figure}

\begin{figure}
\epsscale{0.85}
\plotone{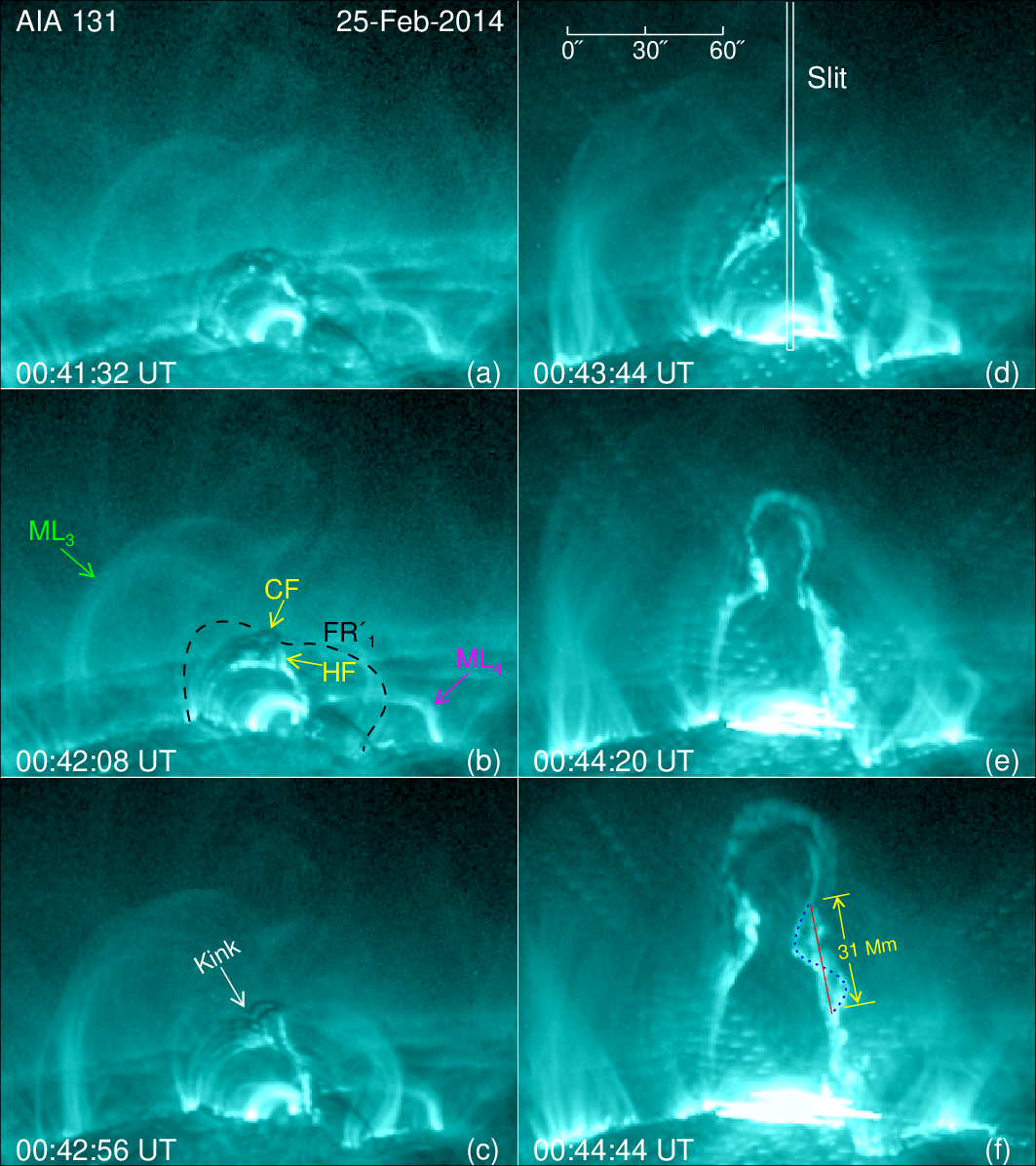}
\caption{AIA 131 \AA\ images showing the eruption of the filament and the associated hot
flux ropes (also see animation~3).
%The black curve in panel (b) displays the general outline of the hot flux rope formed due to T-C reconnection. 
%The purple and blue dotted lines in panel (c) indicate the two branches of ML$_4$.
%The opened narrow box in panel (d) indicates where the time-slit map in Figure~4(a) was made.
The FOV is 200\arcsec $\times$ 150\arcsec.
%The FOV is the same as that of Figure~2.
\label{fig3}}
\end{figure}

\clearpage

\begin{figure}
\epsscale{0.9}
\plotone{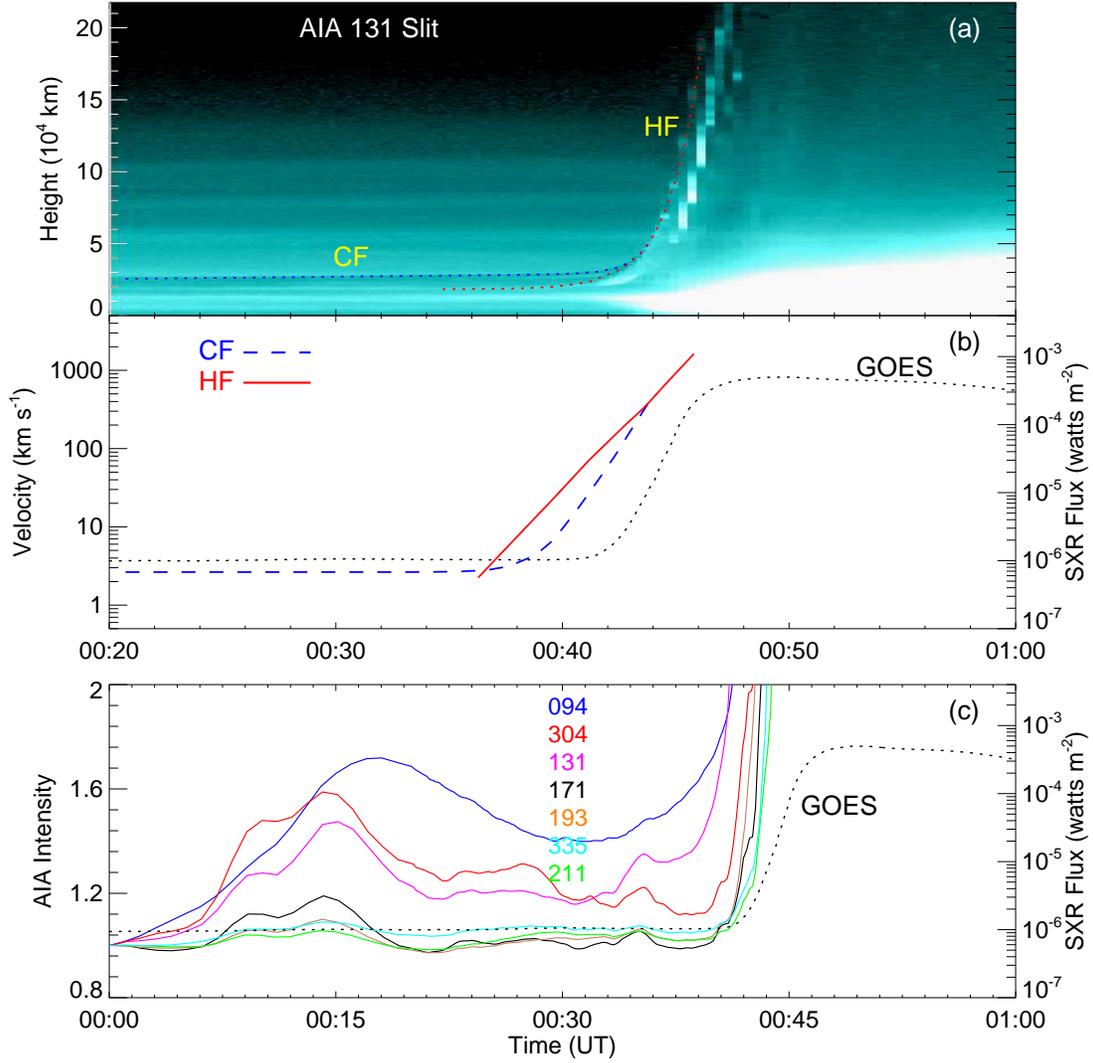}
\caption{(a) Time-slit map from the AIA 131 \AA\ images;
(b) Time variations of the projected velocities of CF and HF;
%({\it plus})
(c) Time profiles of the AIA intensity flux in 94 \AA\ (blue), 304 \AA\ (red), 
131 \AA\ (purple), 171 \AA\ (black), 193 \AA\ (orange), 335 \AA\ (turquoise),
and 211 \AA\ (green) from 00:00 UT to 01:00 UT;
The black dotted curves in panels (b) and (c) are the light curves from the $\it{GOES}$ 1--8 \AA\
channel.
\label{fig4}}
\end{figure}

\begin{figure}
\epsscale{1.0}
\plotone{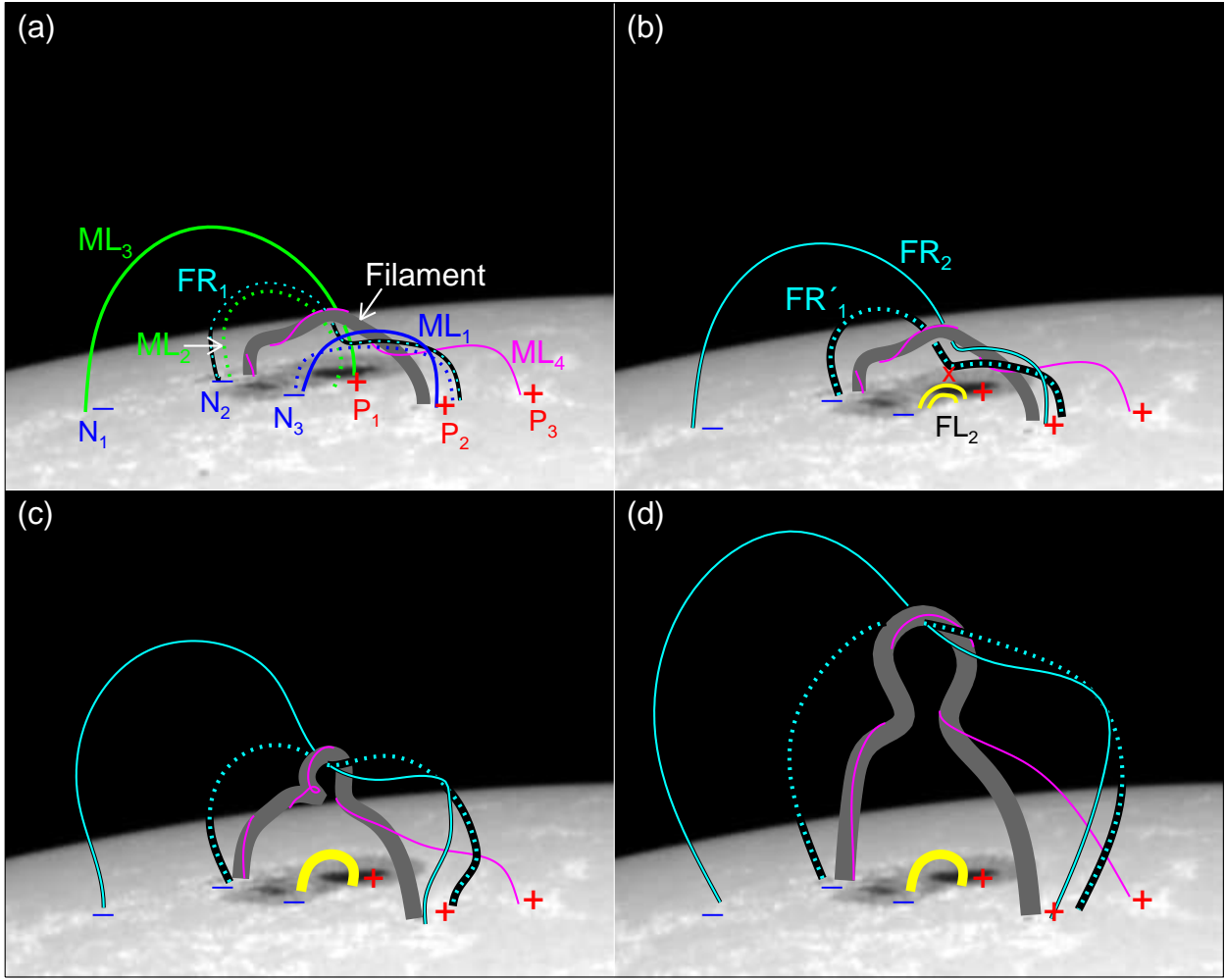}
\caption{Three-dimensional cartoons describing the early phase of the ejective
eruption case. The backgrounds are HMI continuum intensity images.
The FOV is 200\arcsec $\times$ 160\arcsec.
%The blue, green, and purple lines respectively represent ML$_1$, ML$_2$, ML$_3$, and 
%ML$_4$ before the onset of the eruption.
%The plus (+) and minus (-) denote the positive and negative polarity regions.
\label{fig5}}
\end{figure}

\clearpage

%% If you use the table environment, please indicate horizontal rules using
%% Do not put multiple tabular environments within a single table.
%% The optional \label should appear inside the \caption command.

\clearpage

%% If the table is more than one page long, the width of the table can vary
%% from page to page when the default \tablewidth is used, as below.  The
%% individual table widths for each page will be written to the log file; a
%% maximum tablewidth for the table can be computed from these values.
%% The \tablewidth argument can then be reset and the file reprocessed, so
%% that the table is of uniform width throughout. Try getting the widths
%% from the log file and changing the \tablewidth parameter to see how
%% adjusting this value affects table formatting.

%% The \dataset{} macro has also been applied to a few of the objects to
%% show how many observations can be tagged in a table.

%% Tables may also be prepared as separate files. See the accompanying
%% sample file table.tex for an example of an external table file.
%% To include an external file in your main document, use the \input
%% command. Uncomment the line below to include table.tex in this
%% sample file. (Note that you will need to comment out the \documentclass,
%% \begin{document}, and \end{document} commands from table.tex if you want
%% to include it in this document.)


%% The following command ends your manuscript. LaTeX will ignore any text

\end{document}